\documentstyle[12pt]{article}
\begin{document}\baselineskip=18pt
\def\be{\begin{equation}}
\def\ee{\end{equation}}
\vskip 1cm
\begin{tabbing}
\hskip 11.5 cm \= IC/96/51\\
\>hep-th/9604063\\
\>March 1996
\end{tabbing}
\vskip 1cm
\begin{center}
{\Large\bf Screening in Two-dimensional QCD}
\vskip 1.2cm
{\large \bf E. Abdalla$^a$\footnote{Permanent address: Instituto de 
F\'\i sica-USP, C.P. 20516, S. Paulo, Brazil.}, 
R. Mohayaee$^b$ and A. Zadra$^c$}\\
\vskip 0.4cm
{{\it International Centre for Theoretical Physics, Trieste, Italy}\\
\vskip 0.4cm
$^a$ elcio@ictp.trieste.it\\
 $^b$ mohayaee@ictp.trieste.it\\
$^c$ azadra@uspif.if.usp.br}
\end{center}
\abstract 

We discuss the issue of screening and confinement of external colour 
charges in bosonised two-dimensional quantum chromodynamics.
Our computation relies on the static solutions of the semi-classical 
equations of motion. The significance of the different representations of 
the matter field is explicitly studied. We arrive at the conclusion that 
the screening 
phase prevails, even in the presence of a small mass term for the
fermions. To confirm this result further, we outline the 
construction of operators corresponding to screened quarks. 
 
\vfill\eject


\section{Introduction}
\indent

The issue of confinement of fundamental constituents of matter is a 
long-standing problem of theoretical physics whose solution has evaded full 
comprehension up to now. However, significant progress has been made 
towards making a clear distinction between the 
apparently related phenomena of screening and confinement.
In fact, in two-dimensional quantum electrodynamics, the original naive 
definition of 
confinement, as the absence of a pole in the gauge-invariant fermionic
two-point function, has been replaced 
by a more subtle formulation which depends on the introduction of flavour 
quantum numbers \cite{rothe2, elcio1}. In the screening phase, any attempt
to separate a pair of charged particles, bound by a weak 
potential, leads to vacuum polarisation which prevents the
recognition of the individual charges. The 
introduction of the above mentioned flavour quantum numbers overcomes 
this problem: such new quantum numbers are distinguishable in the screening 
case, but not in the confining picture.

In this case \cite{schwinger}, a rich physical structure exists, 
and the issue of confinement is satisfactorily settled \cite{rothe1}. The 
first step towards understanding the problem in
the abelian case is taken by studying the potential created by two opposite
charged particles ($q$,$\bar q$) a distance $L$ apart. 
We determine which of the two phases, {\it i.e.}\ screening or confinement, 
dominates by searching for the solitonic solutions of the classical
field equations. These equations 
are obtained for QED coupled to the external
charges ({\it i.e.}\ charges described by dynamical but classical quarks).
In the solitonic solutions the sum of the non-vanishing external and 
fundamental charges has to be zero \cite{ellis}.
A simple mechanical problem equivalent to this 
setting has been introduced in \cite{rothe1} (see also \cite{elcio1}).
In this picture, the solitonic states are equivalent to the 
configuration in which one of the charged particles stays at the origin
while the other approaches infinity. If the corresponding energy
is finite, we have a single
charge which is screened. If the energy grows linearly 
with distance the particle is confined. The latter phenomenon 
arises in the massive Schwinger model.
 
In two-dimensional quantum chromodynamics the situation is different
and more involved \cite{elcio2}. One of the main
differences concerns the vacuum structure. In the
Schwinger model, the
vacuum structure arises from a unitary operator which acts as a phase
on the physical Hilbert space, characterising the ground
state of the theory, which is infinitely degenerated and is labelled 
by a continuous parameter $\theta$, $0\le\theta < 2\pi$. 
In the massive case, the vacuum parameter 
modifies the dynamics by entering the theory explicitly at the 
lagrangian level. In that case, the issue of
confinement can only be settled after a knowledge of the value of the 
$\theta$ parameter. In general there is a confining term, in the expression 
for the potential, whose scale is dictated by the fermion mass parameter. 
This term disappears
for $\theta=\pi$ and for heavy quarks, leaving only a strong version of 
the screening term. In QCD$_2$, the $\theta$ 
vacuum-parameter  is effectively replaced by a dynamical field 
in the resulting semi-classical equations of motion, and the 
task of distinguishing 
between the screening and the confinement phase is a more subtle one. 

In both theories, however, the question of screening versus confinement 
can be reformulated in terms of the equations of motion of
the electric field and the bosonised matter field. 
The answer then depends on whether the Coulomb interaction survives, or 
whether the Higgs phase dominates \cite{gross}.
 
Our aim in this article is to answer the following questions: 
given two-dimensional QCD 
with an arbitrary number of dynamical fermions coupled to  background 
charges (possibly fractional), are   
these background charges screened or confined? Moreover, if these   
charges are screened, can one define exotic states \cite{elcio1} carrying
flavour quantum numbers? Our results show that the
background charge density is always screened. This result is confirmed by 
our further analysis of the exotic states.

We start by reviewing the Schwinger model, summarising some of the
earlier results in order to set up the
language and outlining the construction of the exotic states.
The procedure will be the 
paradigm of our discussion of two-dimensional QCD. In Section 3,
we study the non-abelian problem. We recall the massless case where the 
system is in a screening phase. Then, we turn to the 
more complicated case of massive two-dimensional QCD. We consider all the 
possible solutions of the classical equations of motion and show 
that the ones which satisfy the requirement of finiteness of 
energy and suitable boundary conditions lead to 
screening. In Section 4, we study QCD with fermions in 
higher representations. We point out how the theory is affected by the 
change of representation and re-establish the results obtained in 
Section 3. In Section 5, we discuss the construction of exotic states 
which correspond to operators that carry flavour quantum numbers but no 
colour charge \cite{rothe2}. Thus, we provide a further evidence for
the screening phase. We discuss our conclusions and further open problems 
in the last section.

\section{The Schwinger model}
\indent

In this section, we review the Schwinger model coupled to a pair of 
background charges $(q,\bar q)$ at positions
$(-{L\over 2} , {L\over 2})$. This introduces an additional constant
electric field in the space between the two charges. We follow
reference \cite{rothe1} and start by writting 
the contribution of the two probe particles to the charge density, {\it i.e.}
\be
J^0=q\left[\delta(x-{L\over 2})-\delta(x+{L\over 
2})\right]=-{e\over\sqrt\pi}{\partial Q\over\partial x},
\ee
where
\be
Q={-q\over e}\sqrt\pi\left[\Theta(x-{L\over 2})-\Theta(x+{L\over 2})\right],
\ee
and $\Theta$ is the step function.
In the bosonised form of the action, or equivalently in the effective 
action, this background density contributes through the anomaly term, 
corresponding to the meson mass term. That is to say, the lagrangian is

\be
{\cal L}={1\over 2}(\partial_\mu E)^2-{e^2\over 2\pi}(E-Q)^2
-2m^2\big[1- {\rm cos}(2\sqrt\pi E+\theta)\big],
\ee
where $E$ is the meson field, $e$ is the
electric charge unit and $m$ is the 
electron mass. The modification of the lagrangian brought about by
the $\theta$ vacuum is transparent
in the mass term. The inter-charge potential energy is 
obtained after finding the field configurations which minimise the 
hamiltonian.
By considering time-independent fields we obtain the hamiltonian
\be
H=\int dx\left\{{1\over 2}{E^\prime}^2 +{e^2\over 
2\pi}(E-Q)^2+2m^2 \left[1-{\rm cos}(2\sqrt\pi E+\theta)\right]\right\}.
\ee
Thus, we may use an analogy with a simple mechanical problem. In this 
setting, $x$ is the time evolution variable, $E$ is the position of a
particle and $H$ is the action. Minimising this action leads to the 
Euler-Lagrange equations. 

The inter-charge potential is defined by the change in energy caused by 
the presence of the probe charges $({\it i.e.}\, \, V(L)=H(L)-H(0))$. 
One can see that this potential is given by
\be
{dV(L)\over dL}=-{e q\over 2\sqrt\pi}
\big[E(L/2)+E(-L/2)\big]+{q^2\over 2},
\ee
where we impose the boundary conditions $E(\pm\infty)=0$ and 
require the continuity of $E$ and its  derivative $E^\prime$ at 
$x=\pm L/2$. 
For massless fermions, it is easy to solve the problem by quadratures and 
one finds a screening potential,
\be
V(L)={q^2\sqrt\pi\over 2e}(1-e^{-eL/ \sqrt\pi}).
\ee
For non-vanishing fermion mass the computation is similar, although the 
integral cannot be computed in closed form. In the approximation for 
heavy fermions, the particle moves in the proximity of the origin, around 
which the solution is expanded. We obtain 
for $\theta=0$ and for small separations
\be
V(L)={e^2 q^2\over 2\pi\alpha^3}(1-e^{-\alpha L})+{q^2\over 2}
(1-{e^2\over\pi\alpha^2})L ,
\ee
where $\alpha=\sqrt{e^2/\pi+8\pi m^2}$. Hence, a confining term
appears in the potential.

Next, we reconsider the hamiltonian (4) in the general case when
the mass $m$ and the $\theta$-parameter are non-vanishing.
We must solve the equation of 
motion of $E(x)$, such that $E$ approaches its vacuum expectation value
at $\pm\infty$. That is to say $E(\pm\infty)=E_\theta$ where the constant 
$E_\theta$
is the solution of the equation which minimises the potential, {\it i.e.},
\be
{\rm sin}(2\sqrt\pi E_\theta+\theta)=-{1\over 4\pi\sqrt\pi}{e^2\over m^2} 
E_\theta.
\ee
For large mass, $m>>e$ , we can solve the equations perturbatively. Defining
$F$ as the deviation of $E$ from the vacuum state,
\be
E=F+E_\theta,
\ee
we find the equation of motion of $F$ to be,
\be
F^{\prime\prime}-\left({e^2\over \pi}+8\pi m^2 {\rm cos}(2\sqrt\pi 
E_\theta +\theta)\right)F=-{e^2\over\pi}Q.
\ee
Finally, using the boundary conditions $F(\pm\infty)=0$, we 
compute the inter-charge potential energy. In the case $q=e$, we find
\be
V(L)\simeq{e^4\over 2\pi\alpha^3}(1-e^{-\alpha L})+{e^2\over 
2\pi}(1-{e^2\over \pi\alpha^2})(\theta-\pi)L .
\ee 
For $\theta=\pi$, the confining term vanishes and  we obtain a purely 
screening potential. In such a case screening plays a rather
dominant r\^ole, being more 
effective than in the massless case.

However, this semi-classical description of the inter-charge energy 
eventually fails, due to pair production. Pair production will
always lead to screening when the external charges are integer
multiples of the fundamental charges. Similarly, in the Wilson loop 
approach, where the perimeter versus area law presumably distinguishes 
between screening and confinement, the presence of dynamical fermions 
implies the perimeter law. Therefore, one needs a more
elaborate formulation.
The question as to whether there is confinement can be definitely
answered if we introduce further quantum numbers into the theory and 
verify whether these new quantum numbers can be measured 
asymptotically (screening) or not (confinement).

Such a task has been accomplished in two-dimensional QED where flavour 
quantum numbers were introduced. If the fermions belong to a flavour 
SU(k) representation (abelian) then bosonisation leads to the hamiltonian

\begin{eqnarray}
H&=&\int({\cal H}_0 +V),\\
{\cal H}_0&=&{1\over 2}\dot E^2+{1\over 2}{E^\prime}^2+
{1\over 2}\sum_{i=1}^{k-1}(\dot\phi^2_i+{\phi_i^\prime}^2),\\
V&=&{1\over 2}{ke^2\over\pi}E^2-2m^2\sum_f
{\rm cos}(2\sqrt{\pi\over k}E+2\sqrt\pi\chi_f+\theta) ,
\end{eqnarray}
where $\chi_f=\sum_i\phi^i\tau^i_{ff}$. For SU(k), $\sum^k_{f=1}\chi_f=0$. 
The 
fields $\phi^i$ are the potentials of the diagonal part of the SU(k) currents
\be
J_\mu^i=-{1\over\sqrt\pi}\epsilon_\mu^\nu\partial^\nu\phi^i ,
\ee
while $J_\mu=-\sqrt{k\over\pi}\epsilon_{\mu\nu}\partial^\nu E$ is 
the U(1) current.

One of the characteristics of  of the Schwinger model with flavour is
the fact that periodicity with respect to $\theta$ changes. Indeed,
instead of being periodic under $\theta\to\theta +2\pi$, the theory is
rather symmetric under $\theta\to\theta +2\pi/k$, where $k$ is the number 
of flavours. One can construct the following 
flavour-carrying fermionic operator \cite{rothe2}
\be
{\cal F}_f(z)=e^{i\sqrt\pi\gamma_5\chi_f+i\sqrt\pi\int_{x^\prime}^\infty dz^1
\dot\chi_f(x^0,z^1)},
\ee
in terms of which mesons ${\cal M}$ and baryons ${\cal B}$ are
\be
{\cal M}={\cal F}_f{\cal F}_f^{\dagger}, \,\,\,\, \,\,\,\,\,\,{\cal 
B}=\Pi_{i=0}^k{\cal F}_{f_i}.
\ee

In the massless case, ${\cal F}_f$ are observable quantities which 
correspond to screened quarks. We verify that they carry SU(k) charge,
\be
\left[Q^i,{\cal F}^\pm_f(x)\right]=-\tau^i_{ff}{\cal F}^\pm_f(x),
\ee
but no U(1) charge.

In the massive Schwinger model the situation changes again, because in 
general ${\cal F}^\pm_f(x)$ does not commute with the mass term. Indeed
\begin{eqnarray}
& &\left[{\cal F}^\pm_f(x),\sum_f{\rm 
cos}\left(2\sqrt{\pi\over k}E(y)+2\sqrt\pi\chi_f(y)+
\theta\right)\right]\nonumber\\
& &\qquad\qquad=\sum_f\left[{\rm cos}\left(2\sqrt{\pi\over 
k}E+2\sqrt\pi\chi_f+\theta-{2\pi\over 
k}\Theta(y^1-x^1)\right)\right.\nonumber\\
& &\left. \qquad\qquad-{\rm cos}\left(2\sqrt{\pi\over k} 
E+2\sqrt\pi\chi_f+\theta\right)\right]{\cal F}_f(x).
\end{eqnarray}
This shows again that ${\cal F}_f$ does not generate an eigenstate
of the Hamiltonian for non-vanishing fermion mass, unless
$\theta =\pi/k$ where one can show that after a dressing of the operator 
${\cal F}_f$ by the kink operator, it commutes with the mass term 
\cite{elcio1}, and we are once more back to the screening situation.

\section{Non-abelian gauge theories}
\indent

Two-dimensional QCD with fermions in an arbitrary representation may be 
described in terms of quark fields $\psi^f_i$ 
where $i$ is the usual colour index and $f=1,\cdots ,k$ is a flavour 
quantum number. The full QCD lagrangian is 
\be
S=\int d^2 x \left[-{1\over 4}{\rm tr}F_{\mu\nu}F^{\mu\nu}+
\bar\psi_i^f(i{\not\!\partial}
\delta^{ij}+e{\not\!\! A}^{ij})\psi^f_j-m^\prime\bar\psi^f\psi^f\right].
\ee

Although this model is two-dimensional, it is by no means simple and
has, so far, evaded an exact solution. For an 
SU(N) gauge group and $k=1$ the large N limit was
studied \cite{thooft} and the approximate 
spectrum was obtained. Bosonisation of the fermionic determinant in 
terms of the Wess-Zumino-Witten functional helps 
understanding part of the structure of the theory \cite{elcio3}. That
is, one can use the following identity: 
\be
\int {\cal D}\bar\psi{\cal D}\psi e^{i\int d^2x 
\bar\psi(i {\not D}-m)\psi}=\int{\cal D}\hat g e^{i\Gamma[\hat 
g,A]+i{m^\prime}^2\int d^2x {\rm tr}(\hat g+{\hat g}^{-1})},
\ee
where the gauged Wess-Zumino-Witten action is
\begin{eqnarray}
\Gamma[\hat g,A]&=&\Gamma[\hat g]+{1\over 4\pi}\int d^2x\, {\rm 
tr}[e^2A^+A_+-e^2A_+\hat gA_-{\hat g}^{-1}\nonumber\\
& &-ieA_+\hat g\partial_-{\hat g}^{-1}-ieA_-{\hat g}^{-1}\partial_+\hat g],
\end{eqnarray}
and
\begin{eqnarray}
\Gamma[g]&=&{1\over 8\pi}\int d^2x\, {\rm 
tr}\partial_\mu  g^{-1}\partial^\mu g\nonumber\\
&+&{1\over 4\pi}{\rm 
tr}\int_0^1 
dr\int d^2x\,\epsilon^{\mu\nu}{\tilde g}^{-1}{d\tilde g\over dr}{\tilde 
g}^{-1}\partial_\mu \tilde g{\tilde g}^{-1}\partial_\nu \tilde g
\end{eqnarray}
is the usual WZW action, and $\tilde g(r,x)$ interpolates 
a trivial configuration $\tilde g(0,x)=1$ to $g(x)$ itself, 
{\it i.e.} $\tilde g(1,x)=g(x)$.
The mass parameter $m$ in the bosonic theory has been chosen with an 
arbitrary renormalisation and is related
(but not necessarily equal) to the 
fermion mass $m^\prime$. The flavour $f$ in 
such a scheme is introduced in the usual manner. We decompose the gauge 
fields into the SU(N)-valued fields $U$ and $V$, {\it i.e.},
\be
A_+={i\over e}U^{-1}\partial_+U,\qquad A_-={i\over e}V\partial_-V^{-1} ,
\ee
such that
\be
\Gamma[\hat g,A]=\Gamma[U\hat g V]-\Gamma[UV]=\Gamma[g]-\Gamma[\tilde\Sigma],
\ee
where $g=U\hat g V , \tilde\Sigma=UV$. The jacobian associated with 
transformations (24) is non-trivial and contributes to the action with a 
term 
$-c_v\Gamma[\tilde\Sigma=UV]$. In the massless theory, the dressed field 
$g_f=U\hat g_fV$ decouples. This theory also contains
the negative metric field $\Sigma$ (which is obtained
from $\tilde\Sigma$ after 
some algebraic manipulations, $\Sigma=\beta\tilde\Sigma$) and the ghosts. 
The  $g$ and $\Sigma$-fields together with the ghost system build up the 
vacuum structure of two-dimensional QCD. The 
self-interaction term of the gauge field strength can be 
simplified: first we add a 
term $-{1\over 2}(E+{1\over 2}F_{+-})^2$ to the lagrangian and then we 
define $\partial_+(UEU^{-1})={i\over 4\pi}\beta^{-1}\partial_+\beta$. (These 
manipulations are described in detail in \cite{elcio2,elcio3}.) Finally, we 
arrive at the effective action
\begin{eqnarray}
S_{{\rm eff}}&=&\sum_f\Gamma[g_f]-(c_v+k)\Gamma[\Sigma]+S_{\rm 
ghosts}+k\Gamma[\beta]\nonumber\\
& &+\int d^2z {\rm tr}\left[{1\over 
2}(\partial_+C_-)^2+i\lambda 
C_-\beta^{-1}\partial_+\beta\right]\nonumber\\
   & &+m^2\int d^2 z{\rm tr}\left[\sum_f
(g_f\Sigma^{-1}\beta+g^{-1}_f\Sigma\beta^{-1})\right],
\end{eqnarray}
with $\lambda={e\over 2\pi}(c_v+k)$. 
The equations of motion read
\begin{eqnarray}
{1\over 4\pi}\partial_+(g_f\partial_-g_f^{-1})&=&m^2(g_f\Sigma^{-1}\beta -
\beta^{-1}\Sigma g^{-1}_f),\quad f=1\cdots k,\nonumber\\
& &\\
-{(c_v+k)\over 4\pi}\partial_+(\Sigma\partial_-\Sigma^{-1})&=&
m^2\sum_{f=1}^k(\Sigma g_f^{-1}\beta^{-1}-\beta g_f\Sigma^{-1}) ,\\
-{k\over 
4\pi}\partial_-(\beta^{-1}\partial_+\beta)&+&i\lambda[\beta^{-1}\partial_+
\beta ,C]+i\lambda\partial_+ 
C \, ,\nonumber\\
&=&m^2\sum_{f=1}^k(g_f\Sigma^{-1}
\beta-\beta^{-1}\Sigma g_f),\\
\partial_+^2 C&=&\lambda(\beta^{-1}i\partial_+\beta).
\end{eqnarray}
We proceed by considering the case of massless fermions in the 
fundamental representation ($k=1$) and compute the inter-quark 
potential. We then introduce a pair of classical colour charges of 
strength $q^a$ and a distance $L$ apart. Such a pair
is introduced in the action (26) by means of the substitution
\be
i(\beta^{-1}\partial_+\beta)^a\longrightarrow 
i(\beta^{-1}\partial_+\beta)^a-{2\pi\over e}q^a\bigg(\delta(x-{L\over 
2})-\delta(x+{L\over 2})\bigg),
\ee
where $a$ is a definite colour index. This adds the following new term to 
the action 
\footnote{This corresponds to minus the same term added to the hamiltonian.}
\be
V(L)=\Delta S=S_q-S=-(c_v+1)
q^a\bigg(C_-^a(L/2)-C_-^a(-L/2)\bigg). 
\ee
The equation of motion for $C^a$ is now replaced by
\be
\partial_+^2C^a=i\lambda(\beta^{-1}\partial_+\beta)^a
-(c_v+1)q^a\bigg(\delta(x-{L\over 2})-\delta(x+{L\over 2})\bigg),
\ee
which implies, upon substitution into the $\beta$-field equation of motion,
\begin{eqnarray}
& &\partial_+\left({i\over 4\pi\lambda}\partial_-\partial_+ C+[\partial_+ 
C,C]+i\lambda C\right)=\nonumber\\
& &\qquad\left({-iq^a\over 2e}\partial_-
+(c_v+1)[q^a,C]\right)
\left[\delta(x-{L\over 2})-\delta(x+{L\over 
2})\right].
\end{eqnarray}
As we are fixing the external charge in the colour space, we
propose the ansatz $C^a=q^a f(x)$, which renders the problem essentially 
abelian. As a result we find the following solution for the potential (32):
\be
V(L)= {(c_v+1)\sqrt\pi\over 2}{q^2\over e}(1-e^{-2\sqrt\pi\lambda L})
\ee
which implies a screening phase.

A more general discussion is, however required. Several points 
have been discarded in the previous discussion. First, we did not take  
account of the vacuum structure. In fact, two dimensional gauge theories may 
have a non-trivial $\theta$-vacuum which arises from charges placed at 
infinity \cite{witten}, or from
further degeneracies related \cite{elcio4} to the BRST structure of the 
massless fields and ghost system \cite{karabali}. In the massless case these 
issues are rather marginal. However, they may play an important r\^ole in 
the massive case where the fields, which previously described only the 
vacuum structure, now effectively couple to the $\beta$-sector. This 
enlarges the physical subspace \cite{elcio5}.

Let us, again, consider the one flavour theory ({\it i.e.}\ the fundamental 
representation for the fermions). Since, in effect, we will be solving the 
equivalent mechanical problem, we may rewrite the theory in terms
of scalars rather than matrix-valued fields, as follows:
\be
g=e^{i2\sqrt\pi\varphi\sigma_2},\qquad \beta=e^{i2\sqrt\pi 
E\sigma_2},\qquad \Sigma=e^{-i2\sqrt\pi\eta\sigma_2},
\ee
where the external charges are taken in the 
direction $\sigma_2$  in the SU(2) 
space\footnote{$\sigma_i$ are the Pauli matrices.}. This parametrisation 
can be simply extended to the case of a general gauge group SU(N). In the 
massive theory, the equations of motion are\footnote{We leave the 
Casimir $c_v$ as a free parameter, since the
expressions corresponding to the Schwinger model will simply be 
obtained from the SU(N) model by taking the limit $c_v\to 0$.}
\begin{eqnarray}
\partial_+\partial_-\varphi&=&-4\sqrt\pi m^2 {\rm 
sin}2\sqrt\pi(E+\varphi+\eta),\\
\partial_+\partial_-\eta&=&{4\sqrt\pi\over c_v+1}m^2 {\rm 
sin}2\sqrt\pi(E+\varphi+\eta),\\ 
\partial_+\partial_- E+4\pi\lambda^2 E&=&-4\sqrt\pi m^2{\rm 
sin}2\sqrt\pi(E+\varphi+\eta)\nonumber\\
& &-2\sqrt\pi(c_v+ 1)\lambda 
q\!\!\left[\Theta(x+{L\over 2})-\Theta(x-{L\over 2})\right].
\end{eqnarray}

We notice the existence 
of the massless combination
\be
\partial_+\partial_-\big(\varphi+(c_v+1)\eta\big)=0 ,
\ee
which is, in fact, a remnant of the vacuum structure  
in the massive case. Moreover, in order to compute the potential
we use the static limit where one discards the time derivatives in the 
above field equations. This is equivalent to considering the effective 
lagrangian
\begin{eqnarray}
{\cal L}&=&{1\over 2}{\varphi^\prime}^2-{1\over 
2}(c_v+1){\eta^\prime}^2+{1\over 2}{E^\prime}^2-2m^2{\rm cos}2\sqrt 
\pi(\varphi+\eta+E)\nonumber\\
& &+2\pi\lambda^2E^2+2\sqrt\pi(c_v+1)\lambda E q\left[\Theta(x+{L\over 
2})-\Theta(x-{L\over 2})\right]\nonumber\\
& &-{(c_v+1)^2q^2\over 
2}\left[\Theta(x+{L\over 2})-\Theta({x-L\over 
2})\right]\nonumber\\ 
&=&{1\over 2}{E^\prime}^2+{1\over 2}{c_v+1\over 
c_v}{\Phi^\prime}^2+2\pi\lambda^2 E^2-2m^2{\rm 
cos}2\sqrt\pi(\Phi+E)\nonumber\\
& &+2\sqrt\pi(c_v+1)\lambda q E\left[\Theta(x+{L\over 2})-\Theta(x-{L\over 
2})\right]\nonumber\\
& &-{(c_v+1)^2 q^2\over 2}\left[\Theta(x+{L\over 
2})-\Theta(x-{L\over 2})\right]-{{\psi^\prime}^2\over 2 c_v},
\end{eqnarray}
where $\Phi=\varphi+\eta$ and 
$\psi=\varphi+(c_v+1)\eta$. This change of
variables is possible because only some of the
contributions will couple to the 
physical variables. In addition, we find instanton-like solutions, 
where the field $E$ asymptotically vanishes as $x\to\pm\infty$ and
makes a lump within the range $|x|<L/2$. Thus the 
cosine term may be expanded. 
Later, we have to confirm that the solution is consistent with such a 
condition.

It is worth remaking that the 
abelian case $(c_v\to 0)$ corresponds to a rather singular limit in the
lagrangian (41) 
where the kinetic term for $\Phi$ as well as for $\psi$ have each a divergent 
coefficient. We will see later that at the level of interacting potential 
this limit is smooth.

In the weak-limit approximation, we expand the cosine term. 
Consequently, we diagonalise the hamiltonian and solve 
the equations of motion \footnote{All the forthcoming computations will 
in general be valid for any compact group. In such cases, the mass term 
can always be expanded in terms of
algebra-valued fields after a convenient parametrisation.}.

The diagonalisation of the quadratic lagrangian leads to 
the expression
\begin{eqnarray}
{\cal L}&=&-{1\over 2 c_v}{\psi^\prime}^2\nonumber\\ 
&+&(1+\epsilon a^2)\left\{{1\over 2}{\chi^\prime}^2_++{1\over 
2}m^2_+\chi_+^2+\lambda Q_+\chi_+\right\}\nonumber\\
&+&{(1+\epsilon a^2)\over a^2}\left\{{1\over 2}{\chi^\prime}^2_-+{1\over 
2}m_-^2\chi_-^2+\lambda Q_-\chi_-\right\},
\end{eqnarray}
where we have found it useful to define the following variables :
\begin{eqnarray}
\chi_+&=&{1\over 1+\epsilon a^2}(E-a\Phi),\\
\chi_-&=&{1\over 1+\epsilon a^2}(\Phi+\epsilon a E)
\end{eqnarray}
and the parameters :
\begin{eqnarray}
\epsilon&=&{c_v\over (c_v+1)},\\
a&=&-{8\pi  m^2\over m_+^2- 16\epsilon  m^2},\\
m_{\pm}^2&=&2\pi\left[\left(\lambda^2+(1+\epsilon)2  m^2\right)
\pm\sqrt{\left(\lambda^2+(1+\epsilon)2m^2\right)^2-8\epsilon\lambda^2  
m^2}\right],\nonumber\\
& &\\
Q_\pm&=&q_\pm \left[\Theta(x-{L\over 2})-\Theta(x+{L\over 2})\right],\\
q_+&=&{2\sqrt\pi(c_v+1)q\over (1+\epsilon a^2)},\\
q_-&=&{2\sqrt\pi\epsilon a (c_v+1) q\over (1+\epsilon
a^2)}.
\end{eqnarray}
Solving the corresponding equations of motion yields:
\be
\chi_\pm={\lambda q_\pm\over m_\pm^2} 
\left\{\begin{array}{ll}
{\rm sinh}(m_\pm {L\over 2})e^{-m_\pm|x|}  &  |x|>{L\over 2}
\\
(1-e^{-m_\pm L/2}{\rm cosh} m_\pm x)  &   |x|<{L\over 2}
\end{array}
\right.
\ee
\noindent
from which we obtain the inter-quark potential energy
\begin{eqnarray}
& &V(L)={(c_v+1)^2q^2\over 2}\nonumber\\
& &\quad\times\left[ 
\!\!\left({4\pi\lambda^2-m_-^2\over 
m_+^2-m_-^2}\right)\!\!\!\left({1-e^{-m_+L}\over m_+}\right)
\!+\!\left({m_+^2-4\pi\lambda^2\over 
m_+^2-m_-^2}\right)\!\!\!\left({1-e^{-m_-L}\over 
m_-}\right)\!\!\right].\nonumber\\ & &
\end{eqnarray}
Thus we find two mass scales given by $m_+$ and $m_-$. Both of these 
lead to screening-type contributions.

Next, we compare the results with those obtained for the 
Schwinger model. In the abelian case, 
the combination of the matter boson $\varphi$ and the negative metric scalar 
$\eta$ gives rise to the $\theta$-angle. That is, the combination 
\be
\Phi\equiv\varphi+\eta=\theta 
\ee
appears in the mass term. When fermions are massless, the electric field
and the matter boson decouple. However, due to a Higgs mechanism, the 
electric field acquires a mass and, therefore, a long-range force does 
not exist. This leads to a pure screening potential. On the other hand,
for massive fermions, the electric field couples to the matter boson.
Yet, $\Phi$ remains massless. The coupling through the mass is the origin 
of the long-range 
force in the massive U(1) case. The potential is confining. 

On the contrary, the expression $(52)$ for the potential indicates the 
absence of a long-range force in the non-abelian theory. This can be 
understood by recalling that there are two independent and relevant 
combinations of the fields $\varphi$ and $\eta$. The massless 
combination, which describes the vacuum constraints, is
\be
\psi\equiv\varphi+(c_v+1)\eta.
\ee
This combination decouples from the electric field (see eqn. (39)). The 
other one, $\Phi\equiv\varphi+\eta$, which is massive, is the combination
that couples to $E$. Therefore, as both $E$ and $\Phi$ are massive, there 
is no long-range force. This is confirmed by our explicit computations.

The abelian potential (7) can also be obtained from (52) by taking the 
limit $c_v\to 0$. In this limit, the mass scale $m_-$ tends to zero
and we recover the confinement term.

It is interesting to examine the behaviour of the screening
potential (52) in extreme limits. 
In the strong coupling regime, 
$\lambda^2>>m^2$, the mass parameter $m_+$ dominates ($m_+>>m_-$) and we 
have 
\be
V(L)_{(m<< e)}\simeq {(c_v+1)^2 q^2\over 
2}\left\{\!\!{(1-e^{-2\sqrt\pi\lambda
L})\over 2\sqrt\pi \lambda}+{\sqrt{\pi\epsilon } m\over
\lambda^2}(1-e^{-2\sqrt{2\pi\epsilon} mL})\!\!\right\}.
\ee
On the other hand, in the weak coupling limit, $m>>e$, we obtain
\be
V(L)_{(m>>e)}\simeq{(c_v+1)q^2\over 
4\sqrt\pi\lambda}\sqrt{{1+\epsilon\over\epsilon}}
\left(1-e^{-2\sqrt{\pi\epsilon/(1+\epsilon)}\lambda L}\right).
\ee
In both regimes, the potential is governed by the parameter $\lambda$,
{\it i.e.} by the coupling constant.

\section{Higher representations}
\indent

So far, we have considered fermions in the fundamental representation.
In this section, we continue our quest for confinement by reformulating
the preceding analysis using fermions in higher representation \cite{gross}. 
This can be done by introducing copies of the fermionic fields, that is
\be
{\cal L}_{{\rm fermi}}=\sum_f\big(\bar\psi_f i{\not\!\! D} \psi_f-m^\prime 
\bar\psi_f\psi_f\big).
\ee
This implies that the bosonised version of theory contains a set of 
fields $g_f$, each of them in the fundamental representation.

Following the procedure of the last section, we fix the external charges 
in colour space, parametrise the fields as in equation (36) and take the 
weak-field and static limit. Finally, we arrive at the following 
diagonalised lagrangian 
\footnote{ Note that now $g_f=e^{i\varphi_f\sigma_2}$,   
$f=1,\cdots , k$.}:

\begin{eqnarray}
{\cal L}&=&\sum_{f=1}^k\left[{1\over 2}(\partial_\mu\zeta_f)^2
-4\pi m^2\zeta_f^2\right]-{1\over 2 c_v}{\psi^\prime}^2\nonumber\\  
&+&k(1+\epsilon a^2)\left\{{1\over 
2}{\chi^\prime}^2_++{1\over
2}m^2_+\chi_+^2+\lambda Q_+\chi_+\right\}\nonumber\\
&+&{k(1+\epsilon a^2)\over a^2}\left\{{1\over 2}{\chi^\prime}^2_-+{1\over
2}m_-^2\chi_-^2+\lambda Q_-\chi_-\right\}\, ,
\end{eqnarray}
where the decoupled fields are
\begin{eqnarray}
\zeta_f&=&(\varphi_f+\eta)-\Phi\qquad ,\qquad\qquad \sum_{f=1}^k\zeta_f=0 ,\\
\Phi&=&{1\over f}\sum_{f=1}^k (\varphi_k+\eta) ,\\
\psi&=&\sum_{f=1}^k \varphi_f +(c_v+k)\eta
\end{eqnarray}
and $\chi_{\pm}$ were defined in (43) and (44). The parameters
are
\begin{eqnarray}
\epsilon&=&{c_v\over (c_v+k)},\\
m_{\pm}^2&=&{2\pi\over k}\left[\left(\lambda^2+(1+\epsilon)2 k m^2\right)
\pm\sqrt{\left(\lambda^2+(1+\epsilon)2km^2\right)^2-8\epsilon k\lambda^2
m^2}\right],\nonumber\\
& &\\  
q_+&=&{2\sqrt\pi(c_v+k)q\over k(1+\epsilon a^2)},\\
q_-&=&{2\sqrt\pi\epsilon a (c_v+k) q\over k(1+\epsilon 
a^2)},
\end{eqnarray}
with $a$ and $Q_\pm$ given by (46) and (48).
The finite-energy solutions to the equations of motion should tend to the 
corresponding vacuum states at $x=\pm\infty$. These vacuum states are 
obtained by extremising the full expression for the potential 
(prior to taking the weak-field limit). This leads to the condition 
$E(\pm\infty)=0$\footnote{Recall that in the abelian case we had many 
options for the field E depending on the value of $\theta$, as in eqn. (8).}
 and two possibilities for the field $\Phi$; $\Phi(\pm\infty)=0$ or 
$\Phi(\pm\infty)=\sqrt\pi$. However, only the first choice yields a 
finite-energy solution, which in turn implies $\chi_{\pm}(\pm\infty)=0$.
The finite-energy solutions obey the boundary condition 
$\chi_{\pm}(\pm\infty)=0$ and the same expressions as (51), but with the 
above parameters. The inter-quark potential is
\begin{eqnarray}
& &V(L)={(c_v+k)^2q^2\over 2k}\nonumber\\
& &\quad\times
\left[\!\!\left({4\pi\lambda^2-km_-^2\over   
m_+^2-m_-^2}\right)\!\!\!\left({1-e^{-m_+L}\over m_+}\right)
\!+\!\left({km_+^2-4\pi\lambda^2\over
m_+^2-m_-^2}\right)\!\!\!\left(1-e^{-m_-L}\over 
m_-\right)\!\!\right]\nonumber\\
& &
\end{eqnarray}

We have verified that further solutions obtained by means of
different boundary conditions do not lead to confinement. To explain this
result, we take up the example of $k=2$  \footnote{The generalisation to 
a generic $k$ is trivial.}. We define $\phi_f=\varphi_f+\eta$. The 
saddle points of the potential correspond to $E=0$ and either $\phi_f=0$ 
or $\phi_f=\sqrt\pi$. We consider one of the $\phi_f$s to be in the 
second vacuum state. For instance, we take $\phi_1(\pm\infty)=0$ and 
$\phi_2(\pm\infty)=\sqrt\pi$. The non-trivial choice for $\phi_2$ 
corresponds to a non-trivial $\theta$ vacuum and is equivalent to a 
change of sign in its mass term ($m^2\rightarrow-m^2$). In the weak-field
limit we obtain the following equations of motion:
\begin{eqnarray}
-E^{\prime\prime}+4\pi\lambda^2 E+8\pi 
m^2(\phi_1-\phi_2)&=&-2\sqrt\pi(c_v+2)\lambda 
q\nonumber\\
\times\left[\Theta(x+{L\over 2})\right. &-&\left. \Theta(x-{L\over 
2})\right],\\ 
-{c_v+1\over c_v}\phi_1^{\prime\prime}-{1\over 
c_v}\phi_2^{\prime\prime}+8\pi m^2(\phi_1+E)&=&0,\\
-{c_v+1\over c_v}\phi^{\prime\prime}_2-{1\over 
c_v}\phi_1^{\prime\prime}-8\pi m^2(\phi_2+E)&=&0.
\end{eqnarray}

Thus, we have a system of coupled differential equations. These lead to  
three different values of the mass-squared parameters which correspond to 
three ``Higgs" masses. One of these is negative and, therefore, leads 
to an oscillatory term. We find no solutions which obey all 
the constraints ({\it i.e.}\ which are asymptotically finite,  continuous
and have continuous derivatives at the positions of the 
quarks). Therefore, screening is the only phase of the theory.

We conclude this section by justifying some of the approximations we have 
used so far. Both in the Schwinger and the non-abelian models we have 
assumed that in expanding the cosine term (cos$2\sqrt\pi(E+\Phi)$), which 
appears in the action (3), one can discard higher order
contributions. The solutions, subsequently, found have the following 
property:
\be
\left\vert 2\sqrt\pi(E+\Phi)(x)\right|\leq {2\pi q\over 
e}\left|{(1+a)(1-\epsilon a)\over(1+\epsilon 
a^2)}\right\vert(e^{-m_-L/2}-e^{-m_+L/2}).
\ee
In the non-abelian case, where both mass scales $m_\pm$ are non-vanishing
the right-hand-side of the above equation tends to zero as $L$ tends to 
infinity.
Therefore, the larger the inter-quark distance, the better our approximation.

The abelian limit is rather different: when $c_v$ approaches $0$, the small 
mass $m_-$ tends to zero and this yields
\be
\left\vert 2\sqrt\pi E(x)\right\vert\leq {2\pi q\over e}{\lambda^2\over 
\lambda^2+2m^2}(1-e^{-\alpha L/2}).
\ee
Hence, in the Schwinger model, the approximations hold only if the 
fermion mass is much larger than the coupling constant. Moreover,
for large separations and nonvanishing fermion mass, 
we have to take into account the confinement term, which blows up for 
large distances.
\section{Exotic states}
\indent

In the preceding sections, we have presented a semi-classical analysis of 
two-dimensional QCD. In order to distinguish between the screening and 
the confinement phases, we have used a dipole-dissociation test. If the
particles are confined, an infinite amount of energy is required to 
isolate them. In this case, as the inter-quark distance increases
pair production occurs which obscures the physical interpretation of the 
results. On the other hand, in the screening phase the amount of 
energy required to dissociate the dipole is finite. Although charge (or 
colour) cannot be seen because of vacuum polarisation, further structures 
( or quantum numbers) can be observed.

In this section, we outline the construction of eigenstates of the 
hamiltonian which carry flavour quantum numbers. These are the analogues 
of the exotic states in the Schwinger model (see Section 2). This 
provides a more elaborate confirmation of the screening phase.

We construct the exotic states by means of the fermionic operator
\cite{elcio1}.
\be
{\cal F}_f(x)=\prod_a 
e^{i\sqrt\pi\phi_f^a(x^0,x^1)/(c_v+k)+
i(c_v+k)\sqrt\pi\int_{-\infty}^{x^1}\dot\phi_f^a (x^0,y^1) 
dy^1} =\prod_a{\cal F}_f^a ,
\ee
where the field $\phi_f^a$ does not carry colour charge. \footnote{ We do 
not expect (72) to be the complete operator which describes flavoured 
physical states. Corrections involving multiple commutators, due to the 
non-abelian character of the theory, can appear.} From the semi-classical 
discussion, the combination $(\varphi^a_f+\eta)$ is the natural candidate 
for the operator $\phi^a_f$. This is because we have chosen symmetric 
boundary conditions $(\phi^a_f(+\infty)=\phi^a(-\infty))$ which imply
that $\phi^a_f$ carries no charge. In the quantum theory \cite{elcio5},
the operator $\phi^a_f=\varphi^a_f+\eta$ appears in the BRST current
\be
J_+=c_+\left(ig^{-1}_f\partial_+ 
g_f-i(c_v+k)\Sigma^{-1}\partial_+\Sigma+{\rm ghosts}\right),
\ee
which is conserved (actually vanishing) and leads to the topological charges
\be
Q=\left(\sum_{f=1}^k\varphi_f+(c_v+k)\eta\right)(t,\infty)-
\left(\sum_{f=1}^k\varphi_f+
(c_v+k)\eta\right)(t,-\infty)+\cdots\,.
\ee
where $\cdots$ stands for the commutator-type corrections.

Next, we argue that the operator (72) commutes with the mass term. We
use the parametrisation
\begin{eqnarray}
g&=&e^{i\varphi^1\sigma^1}e^{i\varphi^2\sigma^2}e^{i\varphi^3\sigma^3},\\
\Sigma&=&e^{-i\eta^1\sigma^1}e^{-i\eta^2\sigma^2}e^{-i\eta^3\sigma^3},
\end{eqnarray}
in the SU(N) model and take the commutator of ${\cal F}_f$ with the mass 
term. This shifts $\varphi^a$ by 
$2\pi(c_v+k)$, and $\eta^a$ by $2\pi$. Since SU(2) is a compact group,
we conclude that ${\cal F}_f$ commutes with the hamiltonian.
This result can be generalised to any SU(N) gauge group.

By comparing expression (72) with the fermionic operator (16), we see that
the field $\eta$ plays a r\^ole similar to that played previously by
the sum $\sum_{i=1}^k \psi_k$ in the abelian theory. Consequently, the 
fields are not constrained in the non-abelian model and enjoy canonical 
commutation relations. Thus, kink dressing might be needed \cite{rothe1},
\cite{rothe2}. In 
addition, the $\theta$-vacuum does not enter the expression for the 
fermionic operator (72) in the non-abelian theory.


\section{Conclusion}
\indent

Having used the semi-classical methods, we conclude that two 
dimensional QCD displays screening even when the fermions are massive.

An important r\^ole in our calculation has been played by the
combination of the bosonised matter and negative metric field ({\it i.e.}\ 
$\Phi=\varphi+\eta$).
Since $\Phi$ is now a dynamical field,
the effect of a possible $\theta$-parameter or of a background BRST 
charge is immaterial. Furthermore, $\Phi$ has no 
(colour) topological number, which is essential for the construction of 
colourless states with flavour degrees of freedom.

In the Schwinger model (as well as in QCD on the SU(N)
torus\cite{brss}) such states are--in some particular $\theta$-worlds--the 
remnants of quark states. The U(1) charge is screened
by the gauge field interaction. Nevertheless, a shadow of its presence is 
left on the multiplicative quantum number of the exotic states. For a very 
refined discussion of the problem of screening and confinement
in QED see ref. \cite{schrfred}.

In two-dimensional QCD, we are faced 
with a similar situation. A well-defined colourless fermionic operator 
can be constructed and thus screening is unavoidable. Presumably,
the disappearance of colour can be understood in terms of the 
spurionisation idea which is equivalent to
superconductive condensation \cite{nielsen, rothe3}. This phenomenon is 
expressed by the non-vanishing expectation value of the dressed fermionic 
field \cite{kurak}. Therefore, we search for locally gauge-invariant fields 
which have 
non-zero expectation values. These gauge-invariant quantities can be 
obtained by dressing the fermionic fields with gauge field flux tubes. 
In our bosonised approach, such  a dressing is done by using the gauge
potentials $U$ and $V$.

Finally a few words of caution.
we have treated the fermion mass perturbatively. For large fermion 
masses, the fields obey sine-Gordon-like equations which deserve  
further studies.

Also, it is known that in the Wilson-loop approach the perimeter versus area 
law criteria may fail to distinguish between 
screening and confinement, due to 
pair production. The decisive test for screening is the 
existence of the exotic states which we have considered.
The ``baryon"  (see Section 5) is the best probe of the 
screening mechanism. Although our construction is 
rather robust and gives the correct results in all known limits, we believe 
it is worthwhile to explicitly construct the exotic states 
in higher representations. Unfortunately, only a few results are available
and these are mainly for massless fields \cite{naculich}.

In addition, it is worth investigating into the discrepancy which
exists between different results concerning the fermion condensates
in QCD with adjoint fermions \cite{smilga}.

Finally, it would be interesting to consider the analogue of our 
analysis in 4 dimensional QCD. If our result persists in 3+1
dimensions it would be crucial to understand   
Seiberg-Witten's monopole condensation in supersymmetric Yang-Mills 
theory, where such a condensation is triggered by a mass parameter 
\cite{seiberg}.

\section{Acknowledgements}
\indent

We thank I. Klebanov for discussions, and K. D. Rothe for the critical 
reading of the manuscript.


\end{document}